# QUASAR-370 hybrid phototube as a prototype of a photodetector for the next generation of deep underwater neutrino telescopes

B. K. Lubsandorzhiev

*Institute for Nuclear Research of the Russian Academy of Sciences*

*pr-t 60th Anniversary of October, 7A, 117312 Moscow, Russia.*

*Kepler Center for Astro and Particle Physics, University of Tuebingen*

*Auf der Morgenstelle 14, D-72076 Tuebingen, Germany*

*Postal address: pr-t 60th Anniversary of October, 7a, 117312 Moscow, Russia; phone: +7-495-1353161; fax: +7-495-1352268;*

*E-mail: lubsand@inr.ac.ru, lubsand@pit.physik.uni-tuebingen.de*

**Abstract**

In this paper we show that QUASAR-370 large area hybrid phototube developed for and successfully used in a number of astroparticle physics experiments, the Lake Baikal deep underwater neutrino experiment among them, could be used as a prototype of a photodetector for the next generation of giant neutrino telescopes.

## 1. Introduction.

Experimental high energy neutrino astrophysics has been very active for the last three decades. Presently it evolves to its mature stage with three operating high energy neutrino telescopes: two deep underwater and one underice telescopes. New projects of giant neutrino telescopes with huge effective volumes are in the offing. One of the most important issues of such projects will be photodetector developments. Up to now there are two competing tendencies in the large sensitive area photodetector development for such application – conventional PMT and hybrid phototube.

Hybrid phototubes with luminescent screens have been developed especially for deep underwater neutrino experiments in the beginning of 1980s. The first such phototube was XP2600 [1, 2] developed by PHILIPS Laboratories (now Photonis Group) for the pioneering deep underwater neutrino detector project DUMAND [3]. For the Lake Baikal neutrino telescope [4], the first deep underwater neutrino telescope in the world, the QUASAR-370 hybrid phototube has been developed [5-7].

Presently there is a renewed interest in hybrid phototubes with luminescent screens [8-9] because of their excellent performance. It was shown in [10] that phototubes, like QUASAR-370, are very close to the ideal photodetector for deep underwater neutrino telescopes.

## 2. QUASAR-370 hybrid phototube
### 2.1. Operational principle and amplitude resolution

Let's describe briefly the operational principle of the phototube. For more details on the phototube's design and performance the readers are referred to [5-7]. The hybrid phototube with a luminescent screen is a combination of an electro-optical preamplifier with a large hemispherical photocathode with more than $2\pi$ viewing angle and a small conventional PMT. The QUASAR-370 phototube's drawing is presented in Fig. 1. Photoelectrons produced by photons impinging on photocathode of the electro-optical preamplifier are accelerated by high accelerating voltage ~25 kV and hit a luminescent screen which is fixed near the center of the phototube's glass bulb. The luminescent screen is a thin layer of a fast, high light yield inorganic scintillator covered by an aluminum foil to suppress light feedback. Light flashes produced by photoelectrons in the scintillator are registered by a small conventional type PMT. As a result one photoelectron from the preamplifier's photocathode engenders typically 20-30 photoelectrons in the small PMT. The luminescent screen and the small PMT's photocathode can be likened to the first stage of photoelectron amplification of conventional PMTs dynode system. Thus the high gain of the phototube's first stage results in an excellent single photoelectron resolution and along with high accelerating voltage allows to keep low the time jitter of the phototube. The QUASAR-370 phototube has a mushroom shaped glass envelope to improve the isochronism of the electron trajectories. The maximum transit time difference of photoelectrons over the photocathode area is less than 1 ns.

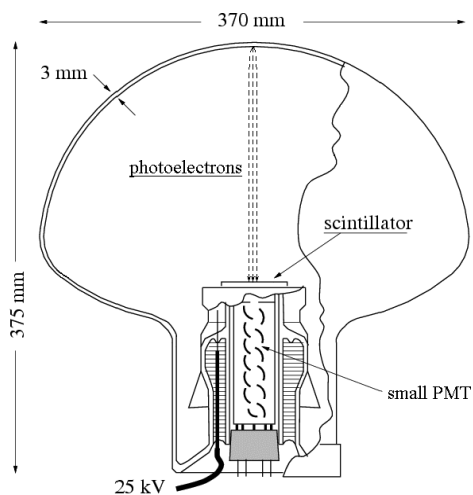

Fig. 1. The QUASAR-370 phototube

The geometry and electron optics of the QUASAR-370 phototube provides very good uniformity of its angular anode sensitivity over the whole photocathode area (see Fig. 2). This parameter of the phototube is very sensitive to mechanical precision of the phototube's assembling. For good tubes the non-uniformity is less than 10%

The single photoelectron resolution of the QUASAR-370 phototube is defined mainly by the gain G of the electro-optical preamplifier. G is the ratio between the number of phototelectrons detected by the small PMT and the number of photoelectrons emitted at the preamplifier photocathode.

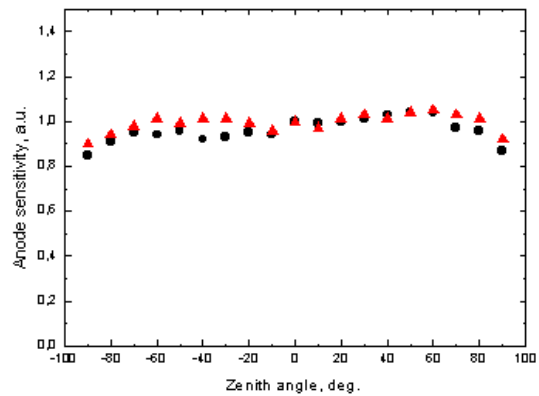

Fig. 2. Dependence of anode angular sensitivity of the QUASAR-370 phototube on Zenith angle in two perpendicular azimuth angles: ▲ - $\varphi = 0$; ● - $\varphi = 90$.

The high amplification factor of the first stage provides very good separation of peaks in the events charge distribution due to one, two, three and even more photoelectrons. We can conclude that the higher the level of the first stage gain G, the better is the single photoelectron resolution. In other words, the higher scintillator light yield Y, the higher is the single photoelectron resolution because the gain G is directly proportional to Y.

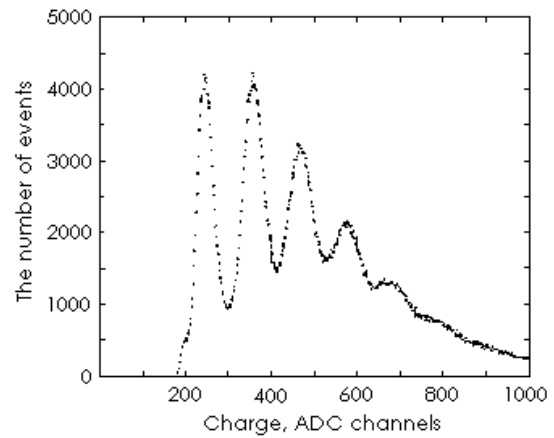

Fig. 3. Charge distribution of multi photoelectron pulses of the QUASAR-370LSO phototube with LSO monocrystal scintillator.

The best single electron resolution of 35% (FWHM) at 25 kV accelerating voltage is reached with the QUASAR-370LSO phototube which is a modification of the QUASAR-370 phototube with LSO monocrystal scintillator. Fig. 3 shows the charge distribution of multi photoelectron events of the QUASAR-370LSO phototube. Very distinct peaks due to up to 7 photoelectrons are clearly seen in this spectrum.

**2.2. Timing**

A single photoelectron pulse of QUASAR-370 is a superposition of G single photoelectron pulses of the small PMT, distributed by an exponential law in time:

$$P(t) \sim \exp(-t/\tau) \qquad (1)$$

with $\tau$ being the decay time constant of the scintillator.

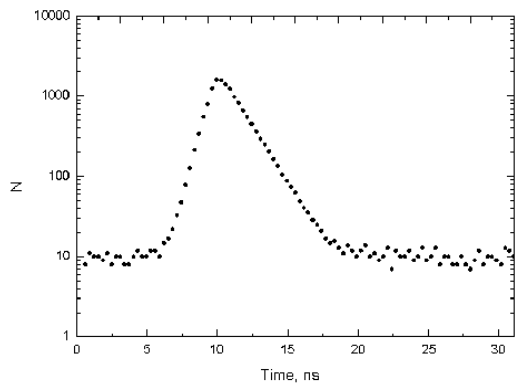

Fig. 4. Single photoelectron transit time distribution of the QUASAR-370YSO phototube with YSO scintillator.

To reach the highest time resolution with such phototubes one should use a double threshold, "fast-slow" discriminator [6]. It consists of two discriminators with different thresholds and integration time constants: a timing, "fast", discriminator with a threshold of $0.25q_1$ and a strobing, "slow", discriminator with a threshold of $0.25Q_1$ ($q_1$ and $Q_1$ are the mean charges of single photoelectron pulses of the small PMT and the electro-optical preamplifier respectively). The events arrival times are defined by the front edge of the first of the G single photoelectron pulses of the small PMT constituting a single photoelectron pulse of the phototube as a whole.

Single photolectron transit time distribution of the QUASAR-370 phototube, Fig. 4, is described well by the following expression:

$$W(t) \sim \exp(-(G/\tau)t) \qquad (2)$$

W(t) is determined by the ratio between the scintillator decay time constant $\tau$ and the first stage gain G. Strictly speaking the equation (2) describes precisely only the right slope of the distribution. The left slope is defined mainly by a scintilation rise time and jitters of the small PMT and electronics. So, the higher the gain G and the shorter decay time $\tau$ of the scintillator, the better is the time resolution of the phototube. Among all the QUASAR-370 phototube's modifications produced over the last 20 years the best time jitter, transit time spread (TTS), of 1 ns (FWHM) at 25 kV accelerating voltage, belongs also to the modification with LSO scintillator – QUASAR-370LSO. The dependencies of TTS (FWHM) for the QUASAR-370 phototube with different scintillators on acceleration voltage is shown in Fig. 5. The TTS was measured for a point-like illumination of the phototube's photocathode in the pole region.

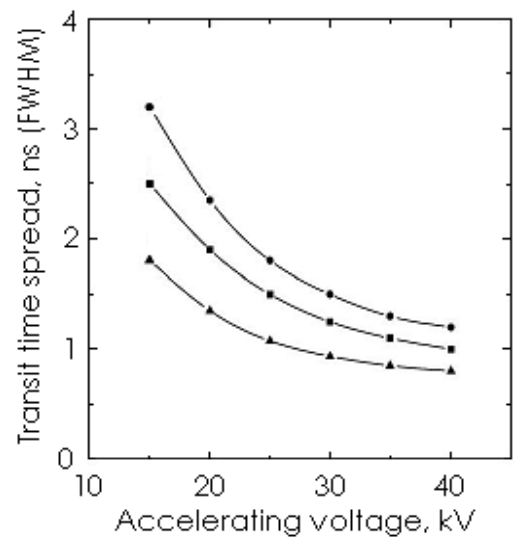

Fig. 5. Dependence of single photoelectron transit time spread (FWHM) of the QUASAR-370 phototube with different scintillators on accelerating voltage. ● - YSO; ■ - SBO and YAP, ▲ - LSO.

The time response of the QUASAR-370 phototube has no prepulses and late pulses as it can be seen in Fig.4.. The reason for the lack of prepulses is the fact that the first stage of the phototube is optically completely isolated from the phototube's photocathode so, photons going through the photcathode don't reach the first dynode of the small PMT. Late pulses are suppressed due to the high gain of the first stage of the phototube. The readers are referred to [11, 12] for more details on prepulses and late pulses.

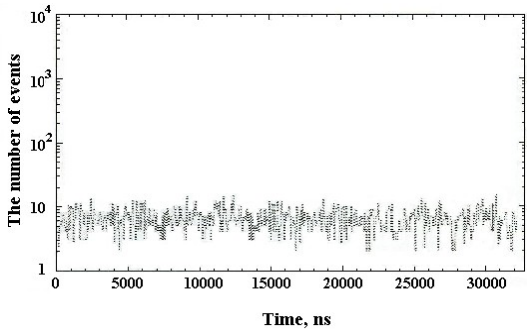

Fig. 6. Time distribution of afterpulses in the QUASAR-370 phototube

Moreover the complete vacuum separation between the electro-optical preamplifier and the small PMT leads to the much lower level of afterpulses in comparison with conventional PMTs. Fig. 6 presents time distribution of afterpulses in the QUASAR-370 phototubes. It should be emphasized that there are no ion peaks as they are observed in conventional PMTs [13]. In fact it is the distribution of just random coincidences due to the phototube's dark current pulses. It is possible to put only an upper limit on the afterpulses probability at the level of 0.1% or even less.

## 2.3. Modifications of QUASAR-370 phototube

For the last 20 years extensive search for the effective and fast scintillator suitable for use in QUASAR-370 phototube have been carried out and a variety of modifications of the phototube based mainly on different scintillators have been developed. The most interesting modifications are equipped with YSO ($Y_2SiO_5$:Ce), YAP (YAlO$_3$:Ce), SBO (ScBO$_3$:Ce), LSO (Lu$_2$SiO$_5$:Ce), etc. So far the best scintillators found for such kind of application are YAP and LSO monocrystal scintillators and SBO phosphor. The best amplitude and timing characteristics are reached with LSO monocrystal scintillator despite of its high nonproportionality. The QUASAR-370 modification with LSO monocrytsal scintillator known as the QUASAR-370LSO has single photoelectron resolution of ~35% (FWHM) and photoelectrons transit time spread of ~1 ns (FWHM), as it is seen in Fig. 3 and Fig. 5.

Table 1

| QUASAR-370 modifications | Scintillator | $R_e$, % (FWHM) | TTS, ns (FWHM) |
|---|---|---|---|
| QUASAR-370YSO | YSO (ph, mc)* | 70-80 | 1.8-2.2 |
| QUASAR-370G | YSO+BaF$_2$ (ph) | 70-80 | 1.8-2.2 |
| QUASAR-370GSO | GSO (ph) | 90 | 2.7-3.0 |
| QUASAR-370YG | YSO+GSO (ph) | 80-90 | 2.2-2.7 |
| QUASAR-370LPO | LPO (mc) | 70-80 | 1.8-2.2 |
| QUASAR-370SBO | SBO (ph) | 40-60 | 1.2-1.5 |
| QUASAR-370YAP | YAP (mc) | 40-60 | 1.2-1.5 |
| QUASAR-370LSO | LSO (mc) | 35 | 1 |

$R_e$ - Single electron resolution; TTS - transit time spread; * - scintillator form: ph – phosphor and mc – monocrystal.

The most interesting modifications of the QUASAR-370 phototube developed so far with a variety of scintillators are listed in Table 1. The scintillator properties play crucial role on the phototube performance [14]. Besides some other modifications like the two-channel QUASAR-370-II and the QUASAR-370D [15] with silicon diode instead of scintillator have been developed in frame of the same glass bulb. It is turned out that the design of the QUASAR-370 phototube allows to make quickly and rather easily modifications and pilot samples of phototubes to test some technical solutions concerning HV, vacuum, scintillator problems etc.

Presently the QUASAR-370 phototubes are used successfully in a number of astroparticle physics experiments. The phototube is the basic photodetector of the Lake Baikal deep underwater neutrino telescope [4, 16] and TUNKA wide-angle extensive air shower Cherenkov detector [7, 17]. More than 15 years operation of QUASAR-370 phototubes in these experiments under very different, harsh conditions proves the phototubes high performances, reliability and robustness.

## 3. Conclusion

The QUASAR-370 hybrid phototube with 37 cm hemispherical photocathode is a unique phototube. It has excellent performance, both timing and amplitude, records paramteres among large area vacuum photodetectors and quite comparable with performances of the best small conventional PMTs: 1 ns (FWHM) time jitter and 35% (FWHM) single electron resolution, ~2π angular acceptance with very good uniformity and immunity to terrestrial magnetic field, the lack of prepulses and late pulses along with suppressed afterpulses rate etc. The phototube is very convenient for modifications. Indeed, the QUASAR-370 phototube as the closest phototube to the ideal photodetector for the next generation giant neutrino projects could be used as a prototype of photodetector for such application. Rich experience accumulated during the phototube's development and operation in a number of astroparticle physics experiments could be very useful in the development of new photodetectors for coming new giant neutrino experiments.

## 4. Acknowledgement

The author is grateful to his colleagues and friends from the Lake Baikal and TUNKA experiments and KATOD Company for many years of very fruitful joint works and Dr. V. Ch. Lubsandorzhieva for careful reading of the paper's manuscript and many useful discussions and invaluable remarks.

**References.**